\documentclass[aps,prl,twocolumn,showpacs]{revtex4-1}
\usepackage{amsmath}
\usepackage[dvips,xdvi]{graphicx}
\usepackage{amssymb}
\usepackage{epsfig}

\begin{document}
\title{Antiferromagnetic spinor condensates in a two-dimensional optical lattice}
\author{L. Zhao}
\author{J. Jiang}
\author{T. Tang}
\author{M. Webb}
\author{Y. Liu}
\email{yingmei.liu@okstate.edu} \affiliation{Department of
Physics, Oklahoma State University, Stillwater, OK 74078}
\date{\today}

\begin{abstract}
We experimentally demonstrate that spin dynamics and the phase
diagram of spinor condensates can be conveniently tuned by a
two-dimensional optical lattice. Spin population oscillations and
a lattice-tuned separatrix in phase space are observed in every
lattice where a substantial superfluid fraction exists. In a
sufficiently deep lattice, we observe a phase transition from a
longitudinal polar phase to a broken-axisymmetry phase in steady
states of lattice-confined spinor condensates. The steady states
are found to depend sigmoidally on the lattice depth and
exponentially on the magnetic field. We also introduce a
phenomenological model that semi-quantitatively describes our data
without adjustable parameters.
\end{abstract}


\maketitle

A spinor Bose-Einstein condensate (BEC) confined in optical
lattices has attracted much attention for its abilities to
systematically study, verify, and optimize condensed matter
models~\cite{StamperKurnRMP,Ueda,spinorLattice}. For instance, it
can quantum simulate the Laughlin-type wavefunctions appearing in
the fractional quantum Hall systems~\cite{lattice3,lattice4}. A
better understanding of these models may directly lead to
engineering revolutionary materials. An optical lattice has been a
versatile technique to enhance interatomic interactions and
control the mobility of atoms~\cite{lattice1,lattice2,fisher89}.
Atoms held in a shallow lattice can tunnel freely among lattice
sites and form a superfluid (SF) phase. The tunneling rate is
exponentially suppressed while the on-site atom-atom interaction
is increased in a deeper lattice. This may result in a transition
from a SF phase to a Mott-insulator (MI) phase at a critical
lattice depth, which has been confirmed in various scalar BEC
systems~\cite{lattice1,lattice2,fisher89,qd}. In contrast to a
scalar BEC, a spinor BEC has unique advantages due to an
additional spin degree of freedom. The SF-MI phase transition is
predicted to be remarkably different in spinor BECs, i.e., the
transition may be first (or second) order around the tip of each
Mott lobe for an even (or odd) occupation number in
lattice-trapped antiferromagnetic spinor
BECs~\cite{StamperKurnRMP,firstorder}.

Spin-mixing dynamics and phase diagrams of spinor BECs in free
space, as a result of spin-dependent interactions and quadratic
Zeeman energy $q_B$, have been well studied with sodium
atoms~\cite{black,faraday,Raman2011,Gerbier2012,JiangBEC,ZhaoUwave,JiangGS}
and rubidium
atoms~\cite{Chapman2005,f2Hirano,f2sengstock1,f2sengstock2}.
Richer spin dynamics are predicted to exist in lattice-trapped
spinor BECs, which allow for a number of immediate applications.
These include constructing a novel quantum-phase-revival
spectroscopy driven by a competition between spin-dependent and
spin-independent interactions, understanding quantum magnetism,
directly detecting spin-dependent three-body and higher-body
interactions, and realizing massive
entanglement~\cite{StamperKurnRMP,spinorLattice,blochLattice10}.
However, dynamics of lattice-trapped spinor BECs have remained
to be less explored, and most of such experimental studies have
been carried out in ferromagnetic $^{87}$Rb spinor BECs
~\cite{SengstockNJP,spinorDL1,spinorDL3,spinorBloch}.

In this paper, we experimentally demonstrate that a
two-dimensional (2D) optical lattice can conveniently tune spin
dynamics and map the phase diagram of $F$=1 antiferromagnetic
spinor BECs. We find that the properties of spinor BECs remain
largely unchanged in the presence of a shallow lattice, while a
sufficiently deep lattice introduces some interesting changes.
First, in every lattice depth $u_L$ which supports a substantial
superfluid fraction, we observe spin population oscillations after
taking spinor BECs out of equilibrium at a fixed $q_B$. These
oscillations are resulted from coherent interconversion among two
$|F=1, m_F =0 \rangle$ atoms, one $|F=1, m_F =+1 \rangle$ atom,
and one $|F=1, m_F =-1 \rangle$ atom. Second, we demonstrate a
lattice-tuned separatrix in phase space and explain it using
lattice-enhanced spin-dependent interactions. Another remarkable
result is our observation of a phase transition from a
longitudinal polar phase to a broken-axisymmetry (BA) phase in
steady states of spinor BECs confined by sufficiently deep
lattices. We find that the steady states depend sigmoidally on
$u_L$ and exponentially on $q_B$. We also introduce a
phenomenological model that semi-quantitatively describes our
experimental data without adjustable parameters. This model takes
into account the observed time evolutions of quantum depletion,
resulting mainly from the lattice-flatten dispersion relation.

We create a BEC of $7\times 10^4$ sodium atoms fully polarized
into the $|F=1,m_F=-1 \rangle$ state in a crossed optical trap via
an all-optical BEC method similar to that of our previous
work~\cite{JiangBEC}. To adiabatically load the BEC into a 2D
lattice, we decompress the optical trap to a value which minimizes
intra-band excitations and ensures approximately constant
Thomas-Fermi radii during a linear ramping of the lattice
potential within $t_{\rm ramp}>40$~ms. We construct the 2D lattice
using two linearly-polarized horizontal beams which originate from
a single-mode laser at $\lambda_L=1064$~nm, have a waist of
$\sim$90$~\mu$m at the condensate, and are retro-reflected to form
standing waves. To eliminate cross interference between different
beams, the two lattice beams are frequency-shifted by 20~MHz with
respect to each other. $u_L$ are calibrated using Kapitza-Dirac
diffraction patterns. Note that all lattice depths studied in this
paper are kept below $15.0(5)E_R$ to avoid SF-MI phase transitions
and thus maintain a sufficient superfluid fraction in our system.
Here $E_R=h^2 k_L^2/(8 \pi^2 M)$ is recoil energy, $k_L=2 \pi /
\lambda_L$ is the lattice wave-number, $M$ is the atomic mass, and
$h$ is the Planck constant. We apply a resonant rf-pulse of a
proper amplitude and duration to lattice-trapped BECs for
preparing an initial state with any desired combination of the
three $m_F$ states at $q_B/h=42$~Hz. $q_B$ is then quenched to a
desired value within a wide range (i.e., $20~{\rm Hz}\leq
q_B/h\leq 1700~{\rm Hz}$). After holding atoms for various amounts
of time $t_{\rm hold}$, we abruptly switch off all lattice and
trapping potentials, and then measure populations of multiple spin
states with standard Stern-Gerlach absorption imaging.

\begin{figure}[t]
\includegraphics[width=85mm]{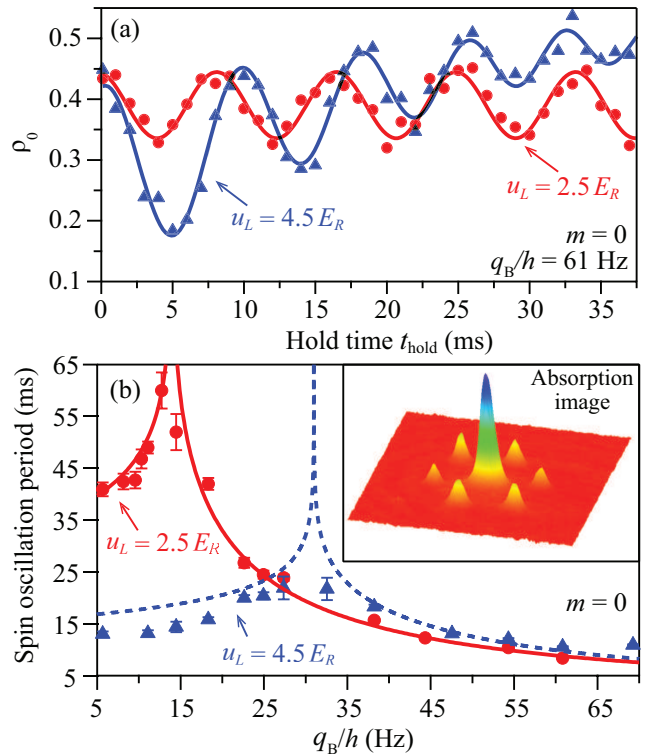}
\caption{(Color online) (a) Time evolutions of $\rho_0$ when $u_L$
equals $4.5E_R$ (blue triangles) and $2.5E_R$ (red circles). Solid
lines are sinusoidal fits to extract oscillation periods. (b)
Oscillation period as a function of $q_{B}$ when $u_L$ equals
$4.5E_R$ (triangles) and $2.5E_R$ (circles). Lines are fits based
on SMA. Inset: an absorption image averaged from 30 raw images.}
\label{shorttime}
\end{figure}

In the presence of a shallow lattice of $u_L<5E_R$, we observe
spin population oscillations which are very similar to those occur
in free space: the oscillations are harmonic except near a
separatrix in phase space where the oscillation period diverges,
as shown in Fig.~\ref{shorttime}. We define $\rho_{\rm m_F}$ as
the fractional population of each $m_F$ state. The total
magnetization $m=\rho_{+1}-\rho_{-1}$ is found to be conserved in
every time evolution studied in this paper. As the lattice is made
deeper, the oscillations appear to damp out more quickly and the
position of the separatrix in phase space shifts to a much higher
$q_B$. Similar to Refs.~\cite{StamperKurnRMP,spinorLattice}, we
apply the Bose-Hubbard model to understand our system. There are
three important terms in this model: the spin-{\it dependent}
interaction energy $U_2$, the spin-{\it independent} interaction
$U_0$, and the tunnelling energy $J$ among adjacent lattice sites.
$U_2$ is proportional to the atomic density in each lattice site,
and is positive (or negative) in $F$=1 $^{23}$Na (or $^{87}$Rb)
BECs. In fact, $U_2/U_0\simeq0.036$ for our $^{23}$Na
system~\cite{spinorLattice}. For the initial state studied in
Fig.~\ref{shorttime}, we find $U_2\simeq q_B$ at each separatrix
in phase space. The observed lattice-tuned separatrix in phase
space (i.e., the separatrix position shifts with $u_L$) is thus
mainly due to the fact that $U_2$ greatly increases with $u_L$.
Fig.~\ref{shorttime}(b) shows a good numerical example: $U_2/h$ is
more than doubled (increased from 14~Hz to 32~Hz) by changing
$u_L$ from $2.5E_R$ to $4.5E_R$. Interestingly, we find that our
data taken at $u_L<5E_R$ can also be fit by predictions derived
from the single-spatial mode approximation (SMA), as shown in
Fig.~\ref{shorttime}(b). SMA assumes that all spin states share
the same spatial wavefunction~\cite{You2005}. Sharp interference
peaks are observed after we release spinor BECs from a shallow
lattice, which indicates coherence and superfluid behavior in the
system. The inset in Fig.~\ref{shorttime}(b) shows a typical
absorption image taken after a 5-ms time of flight (TOF).
\begin{figure}[t]
\includegraphics[width=85mm]{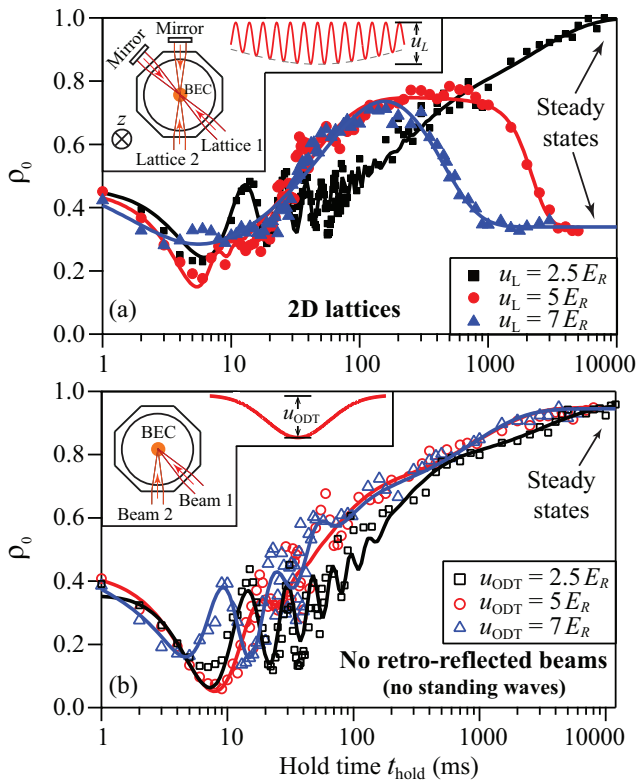}
\caption{(Color online) (a) Time evolution of $\rho_0$ at
$q_{B}/h=42~\rm{Hz}$ and $m=0$ when $u_L$ equals $2.5E_R$ (black
squares), $5E_R$ (red circles), and $7E_R$ (blue triangles).
Inset: a schematic of our lattice setup and an illustration of the
resulting lattice potential. Lines are fits to guide the eye. (b)
Similar to Panel(a) except that each beam is not retro-reflected.}
\label{longtime}
\end{figure}
\begin{figure}[t]
\includegraphics[width=85mm]{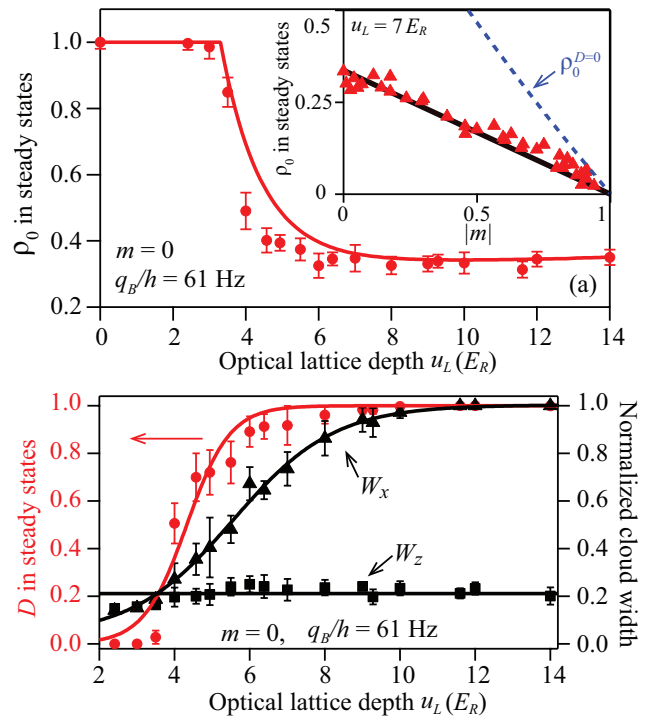}
\caption{(Color online) (a) $\rho_0$ in steady states as a
function of $u_L$ at $m=0$ (main figure), and as a function of
$|m|$ at $u_L = 7E_R$ (inset figure). Solid lines are predictions
derived from Eq.~(\ref{Eqn:qd}). The blue dashed line represents
$\rho_0^{D=0}$ (see Ref.~\cite{rhoMF}). (b) The values of $W_x$
(black triangles), $W_z$ (black squares), and $D$ (red circles) in
steady states as a function of $u_L$. The widths are normalized by
$k_L$. Lines are fits to guide the eye.} \label{QD}
\end{figure}
\begin{figure}[t]
\includegraphics[width=85mm]{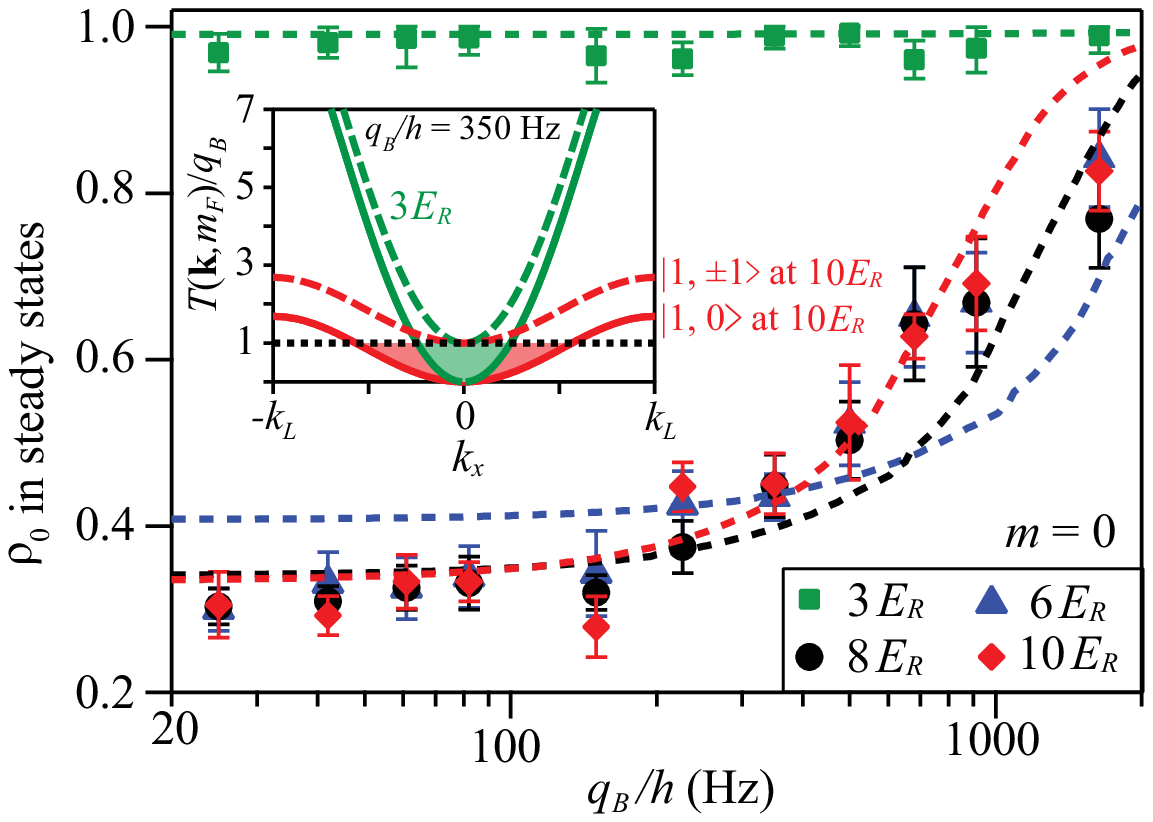}
\caption{(Color online) $\rho_0$ in steady states as a function of
$q_{B}$ at $m=0$ when $u_L$ equals $3E_R$ (squares), $6E_R$
(triangles), $8E_R$ (circles), and $10E_R$ (diamonds). Dashed
lines are predictions derived from Eq.~(\ref{Eqn:qd}). Inset: the
dispersion relations normalized by $q_B$ as a function of $k_x$
when $k_y=k_z=0$. The solid (or dashed) lines represent results of
the $m_F=0$ (or $m_F=\pm 1$) states. The dotted line marks
$T(\mathbf k,m_F)/q_B=1$ and colored regions mark the Region-1
(see text). In the main and inset figures, the green, blue, black,
and red colors respectively represent results at $u_L$ being
$3E_R$, $6E_R$, $8E_R$, and $10E_R$.} \label{steadystates}
\end{figure}

Spin oscillations completely damp out and spinor BECs reach their
steady states when $t_{\rm hold}$ is long enough, as shown in
Fig.~\ref{longtime}(a). Sufficiently deep lattices are found to
bring some interesting changes to the steady states.
Figure~\ref{longtime}(a) demonstrates one of such changes: once
$u_L$ is larger than a critical value, the steady states undergo a
phase transition from a longitudinal polar phase (where
$\rho_0=1$) to a BA phase (where $0<\rho_0<1$) at $m=0$. We repeat
the same measurements with only one parameter changed, i.e., by
blocking the retro-reflected path of each lattice beam. The two
lattice beams are effectively independent, blocking their
retro-reflected paths thus eliminates standing waves and
constructs a crossed optical dipole trap (ODT). Its resulting trap
depth is $u_{\rm ODT}$, as illustrated in the inset in
Fig.~\ref{longtime}(b). Note that the power of every beam in
Fig.~\ref{longtime}(b) is four times of that in
Fig.~\ref{longtime}(a) to ensure $u_L=u_{\rm ODT}$. Our data in
Fig.~\ref{longtime}(b) show that spinor BECs at $m=0$ always reach
the polar phase when there are no standing waves. The dramatically
different results shown in Figs.~\ref{longtime}(a)
and~\ref{longtime}(b) provide strong evidence of the necessity to
understand this polar-BA phase transition with lattice-modified
band structures.

We then study spin oscillations and steady states within a much
wider range of $u_L$ and $m$. Steady states appear to depend
sigmoidally on $u_L$ at a fixed $q_B$, as shown in
Fig.~\ref{QD}(a). The inset in Fig.~\ref{QD}(a) demonstrates
another surprising result: the observed relationship between
$\rho_0$ and $m$ in steady states at a sufficiently large $u_L$
can be well fit by $\rho_0=(1-|m|)/3$, which is drastically
different from a well-known mean-field prediction (i.e.,
$\rho_0^{D=0}$ as illustrated by the blue dashed
line)~\cite{rhoMF}. This mean-field prediction assumes quantum
depletion $D$ is zero, where $D$ is defined as the fraction of
atoms stay in non-zero momentum states. The $D\approx 0$
assumption is correct in free space and in very shallow lattices
for our system, as predicted by Bogoliubov theory~\cite{qd}. We
extract $D$ from TOF images, and confirm $D<5\%$ at $u_L\leq
3E_R$. Note that the trapping frequency in each lattice site is
much bigger than $U_0/h$~\cite{qd}. Our TOF images thus reflect
the momentum distribution at the instant of the lattice release
and enable us to directly measure $D$.

We also find that $D$ increases with $t_{\rm hold}$ and $u_L$, and
approaches one in steady states of spinor BECs when $u_L>10E_R$,
as shown in Fig.~\ref{QD}(b). This lattice-enhanced quantum
depletion is resulted mainly from the lattice-flatten dispersion
relation and lattice-enhanced interactions, and was also observed
in a scalar BEC system~\cite{qd}. We develop one phenomenological
model to take into account the observed $D$. Surprisingly, this
model semi-quantitatively describes our data without adjustable
parameters, as shown in Fig.~\ref{QD}(a) and~\ref{steadystates}.
In this model, the steady states are determined by a comparison
between $T(\mathbf k,m_F=0)$ and $T(0,m_F=\pm 1)$, where
$T(\mathbf k,m_F)$ is the dispersion relation of the $m_F$ state
and $\mathbf k$ is the atom's quasi-momentum. The inset in
Fig.~\ref{steadystates} illustrates two such comparisons in a
shallow lattice ($u_L=3E_R$) and a deep lattice ($u_L=10E_R$).
Note that only the first Brillouin zone is considered, since the
population in higher bands is negligible. Similar to
Ref.~\cite{qd,fisher89,lattice2}, we calculate the dispersion
relation of spinor BECs in a 2D lattice using a Wannier density
function along each of the two horizontal directions with lattices
(the $x$-axis and $y$-axis) and a uniform density function along
the vertical direction without a lattice (the $z$-axis) as
follows,
\begin{equation}
T({\mathbf k},m_F)=4 J \sum_{\alpha=x,y}\sin^2 \left ( \frac{\pi k_\alpha}{2
k_L}\right )+
 E_R \frac{k_z^2}{k_L^2}+q_B m_F^2 ~. \label{Tk}
\end{equation}
Here the linear Zeeman effect is ignored because it remains the
same in the coherent inter-conversions. $T(\mathbf k, m_F=\pm 1)$
is thus shifted up by $q_B$ with respect to $T(\mathbf k, m_F=0)$
at a fixed $u_L$ and a given magnetic field.  The inset in
Fig.~\ref{steadystates} shows that the dispersion relations are
significantly flattened when $u_L$ increases from $3E_R$ to
$10E_R$. In fact, the predicted width of the first band is $\sim 4
J$, where $J$ exponentially reduces with $u_L$~\cite{fisher89,qd}.

We divide $T(\mathbf k, m_F=0)$ into two regions based on
$T(0,m_F=\pm 1)$, i.e., the minimum energy of the $m_F=\pm 1$
states. Region-1 is the first region in which $T(\mathbf k,m_F=0)<
T(0,m_F=\pm 1)$, as illustrated by the colored regions in the
inset in Fig.~\ref{steadystates}. To clearly explain our model
using the dispersion relations shown in Fig.~\ref{steadystates},
we only consider $m=0$ in this paragraph. In Region-1, atoms in
the $m_F=0$ state always have smaller energy than those in the
$m_F=\pm 1$ states. The steady states should be the $m_F=0$ state,
i.e., $\rho_0=1$ which is identical to the mean-field prediction
$\rho_0^{D=0}$. When quantum depletion $D$ is big enough, atoms
start to occupy Region-2 where $T(\mathbf k,m_F=0)\geq T(0,m_F=\pm
1)$. In Region-2 the three $m_F$ states have the same minimum
energy, $T(0,m_F=\pm 1)$, atoms in steady states should thus
evenly distributed among the three states. In other words,
$\rho_0=1/3$ in Region-2, which is identical to the
phenomenological relationship extracted from Fig.~\ref{QD}(a).

We can apply a similar discussion and our model to all non-zero
$m$. Thus $\rho_0$ in the steady states is expressed as,
\begin{align}
\rho_0=\int_{\rm {Region-1}} n(\mathbf k)\rho_0^{D=0}\mathrm d
\mathbf k +\int_{\rm {Region-2}} n(\mathbf k)\frac{1-|m|}{3}
\mathrm d \mathbf k~. \label{Eqn:qd}
\end{align}
Here $n(\mathbf k)$, the normalized atomic density in steady
states, is calculated as follows: $n(\mathbf k)=(1-D)
\delta_\mathbf k+D
\exp[-(k_x^2/W_x^2+k_y^2/W_y^2+k_z^2/W_z^2)/2]$, where $\mathbf W$
is the half-width of a Gaussian fit to a TOF distribution,
$W_y=W_x$, and $\delta$ is a Dirac-delta function.
Figure~\ref{QD}(b) shows that $W_x$ and $D$ sigmoidally increase
with $u_L$, and saturate at their peak values when $u_L>10E_R$. In
other words, atoms occupy all available states and quantum
depletion saturates the first Brillouin zone in a deep lattice. In
contrast, $W_z$ appears to be independent of $u_L$, which implies
the system temperature remains unchanged.

The results derived from our model (Eq.~(\ref{Eqn:qd})) for
various experimental conditions are summarized in
Figs.~\ref{QD}(a) and~\ref{steadystates}. The observed exponential
dependence of steady states on $q_B$ and the sigmodial dependence
on $u_L$ can be explained by this model. In fact, quantitative
agreements between our model and data are found everywhere except
in very high magnetic fields where $q_B/h>1000$~Hz, and in a
moderate lattice depth ($4E_R\leq u_L\leq 6E_R$). The small
discrepancy may be due to the limited resolution in TOF images,
and the resulting larger uncertainties in measuring $D$ and
$\mathbf W$ when quantum depletion does not saturate the first
Brillouin zone. Heating induced by an additional magnetic coil in
creating the very high $q_B$ may also contribute to the
discrepancy.

In conclusion, we have conducted the first experimental study on
dynamics of lattice-trapped antiferromagnetic spinor BECs. Spin
population oscillations and a lattice-tuned separatrix in phase
space have been observed in every lattice where a substantial
superfluid fraction exists. We have found that steady states of
lattice-confined spinor BECs depend sigmoidally on $u_L$ and
exponentially on $q_B$, and undergo a polar-BA phase transition in
a sufficiently deep lattice. We have also developed a
phenomenological model that describes our data without adjustable
parameters. While the underlying physics requires further study,
this paper presents a few thought-provoking results on
lattice-confined spinor BECs.

We thank the Army Research Office and the National Science
Foundation for financial support.


\begin{thebibliography}{24}
\bibitem{rhoMF} Based on the mean-field theory~\cite{JiangGS}, $\rho_0^{D=0}$ in antiferromagnetic spinor
BECs at a fixed $q_B$ equals zero if $q_{B} < U_2(1- \sqrt{1-m^2})$; or equals one if $m=0$; or is the root of the following equation, $1-2 \rho_0^{D=0}-[(1-2\rho_0^{D=0})(1-\rho_0^{D=0})-m^2]/[\sqrt{(1-\rho_0^{D=0})^2-m^2}]=q_{B}/U_2$.
\end{thebibliography}
\end{document}